# The prebiotic pathway from P-bearing iron meteorites to phosphates by DFT modeling

*Stefano Pantaleone,\*[1] Marta Corno,[1] Albert Rimola,[2] Nadia Balucani,[3,4,5] and Piero Ugliengo\*[1]*

[1]Dipartimento di Chimica and Nanostructured Interfaces and Surfaces (NIS) Centre, Università degli Studi di Torino, via P. Giuria 7, I-10125, Torino, Italy

[2]Departament de Química, Universitat Autònoma de Barcelona, 08193 Bellaterra, Catalonia, Spain

[3]Dipartimento di Chimica, Biologia e Biotecnologie, Università degli Studi di Perugia, Via Elce di Sotto 8, I-06123 Perugia, Italy

[4]Osservatorio Astrofisico di Arcetri, Largo E. Fermi 5, I-50125 Firenze, Italy

[5]Université Grenoble Alpes, CNRS, Institut de Planétologie et d'Astrophysique de Grenoble (IPAG), F-38000 Grenoble, France

AUTHOR INFORMATION

**Corresponding Author**

*E-mail: stefano.pantaleone@unito.it, piero.ugliengo@unito.it

ABSTRACT

Among the biogenic macroelements, phosphorus is the one bringing the most fascinating and unsolved mysteries for what concern its prebiotic history. It possibly landed on Earth as a metal phosphide (Schreibersite, $(Fe,Ni)_3P$), throughout the Heavy Meteor Bombardment during the Archean Era. Its subsequent corrosion by water led to P-oxygenated compounds, which is the subject of this kinetic computational study, thus complementing our previous thermodynamic characterization. The reaction was studied at periodic DFT level, simulating the water corrosion reaction on the reactive $Fe_2NiP$ Schreibersite (001)2 surface. Results show that the timescale of the reaction at 350 K is of few hours.

The prebiotic origin of phosphorus (P) is a relatively recent matter of debate in the scientific community in comparison to other fundamental bricks of life, such as amino acids and nucleobases, whose milestone experiments performed by Miller and Oró laid the foundations of modern prebiotic chemistry in the second half of the 20$^{th}$ century.[1,2] As a matter of fact, in 1955 Gulick proposed that P could have come from iron meteorites, specifically from the Schreibersite (Fe,Ni)$_3$P mineral,[3] where all P atoms are in a reduced state and, supposedly, prone to be oxidized due to water weathering in the early Earth. However, this hypothesis remained speculative until the first corrosion experiment on Schreibersite carried out in 2005 by Pasek and coworkers, where several P-oxygenated compounds were produced.[4] More recent experiments demonstrated the capability of Schreibersite to activate other key prebiotic processes; for instance, the formose reaction and the phosphorylation of other fundamental bricks of life, like sugars and nucleobases.[5,6] As a results, among all the solutions proposed to promote phosphorylation (i.e., use of alternative solvents with respect to water, condensing agents, polyphosphates[7–10]), Schreibersite has become one of the most promising minerals to explain the abundance of P in living organisms, considering its income either at the last stages of the Earth formation throughout the meteor bombardment[8,11–13] or its continuous production on the formed Earth through lightning strikes up to the present days.[14] Although many experiments have been carried out on this material, the atomistic mechanism of the water corrosion process leading to the production of P-oxygenated compounds, and even to their inclusion into other prebiotic molecules, is still elusive, and very few computational studies have appeared in the literature.[15–19] Other studies focused on the phase diagrams of different polymorphs of the Fe$_x$P$_y$ binary system, taking into account different combinations of chemical composition, temperature and pressure, to predict the structure of this material under different astrophysical conditions.[20–22] Therefore, in the present paper we address the water corrosion process of the reactive Fe$_2$NiP Schreibersite (001)2 surface by means of periodic DFT simulations, focusing on the formation of the most interesting P-oxygenated compounds, i.e., the phosphite and the phosphate. The (001) surface has been chosen because it is the most reactive one that has a non-negligible contribution to the Wulff shape, as computed in our previous paper.[15] Therefore, it can be thought as an intermediate reactive model between the most stable (110) surface, and either real nanoparticles, containing highly reactive sites such as edges and corners, or defective surfaces where the corrosion already occurs (which therefore present metal and phosphorus vacancies and hydroxylated/hydrogenated species).

Figure 1 shows the reaction pathway of successive water additions (from 1 up to 3 water molecules) to the Schreibersite surface. As already discussed in a previous paper, water cannot directly interact with P in its molecular form.[16] Consequently, each water addition starts on a metal atom (preferentially Fe (Reag-H$_2$O), according to previous calculations)[17] followed by its deprotonation (TS-OH-H, $\Delta G^\ddagger_{298}$ = 35 kJ/mol) to form a metal hydroxide (INT-OH-H), which is an exergonic process ($\Delta G_{298}$ = -51 kJ/mol). At this point, the OH can migrate towards P (TS-POH, $\Delta G^\ddagger_{298}$ = 86 kJ/mol) to form a P-OH moiety (INT-POH), which, on the contrary with respect to water deprotonation, is a slightly endergonic process ($\Delta G_{298}$ = 3 kJ/mol with respect to Reag-H$_2$O). When adding the other water molecules (Reag-2H$_2$O and Reag-3H$_2$O) the reactions proceed in a similar way, *i.e.* water deprotonation followed by the migration of the hydroxyl group. The highest reaction barrier is 95 kJ/mol, which represents the final landing to the phosphorous acid (Prod-P(OH)$_3$) adsorbed on the surface, which overall results in a moderately endergonic reaction ($\Delta G_{298}$ = 9 kJ/mol).

From Prod-P(OH)$_3$ the reaction can bifurcate: on one side, the phosphorous acid tautomerization (P(OH)$_3$), *i.e.* the formation of phosphonic acid (H$_3$PO$_3$), the favorite species in water ($K_{eq}$ = $10^{10.3}$, at room temperature);[23] on the other side, the addition of one more water molecule to obtain the phosphate (see Figure 2). The reaction proceeds with the deprotonation of Prod-P(OH)$_3$, as it presenting a very small barrier ($\Delta G^{\ddagger}_{298}$ = 5 kJ/mol) and a much more stable product (P(=O)(OH)$_2$, $\Delta G_{298}$ = -85.2 kJ/mol, $\Delta G_{298}$ = -94 kJ/mol with respect to Prod-P(OH)$_3$). At this point, we have simulated the adsorption of the fourth water molecule and, as in the other cases, its deprotonation to prepare the following OH attack to the P atom. This deprotonation is endergonic in contrast to previous H$_2$O dissociations ($\Delta G_{298}$ = 20 kJ/mol). However, this is not unexpected as we have already seen in previous papers that the higher the water loading, the less favorable the dissociation reaction, which becomes endergonic for the third and the fourth water molecule.[17] This possibly indicates that, when the corrosion process is in an advanced stage, the metallic and P redox species must dissolve into the water medium in order to generate other surface sites available to further corrosion. The following step is the migration of the OH group from the surface to the P atom with a barrier of 115 kJ/mol, slightly higher with respect to the other ones, thus forming the phosphoric acid (Prod-H$_3$PO$_4$) as final product, which results in an overall exergonic process ($\Delta G_{298}$ = -69 kJ/mol). According to the classical transition state theory, the highest barrier (115 kJ/mol), which represents the rate determining step of the global process, corresponds to a half-life time of about $t_{1/2}$=160-170 days at room temperature (see Eqs. 1 and 2), which is reasonably fast considering the geological timescales covering the prebiotic context. It is worth mentioning that the room temperature was taken as reference; however, it is possible that the temperature of meteorites easily exceeds this mild parameter, in particular considering the extreme post-shock conditions after a meteor impact, where temperatures may reach 2000 K, which should further speed up the reaction.[24–26] As a matter of fact, a temperature of 350 K is enough to speed up the reaction to $t_{1/2}$ of few hours.

In conclusion, this paper reports a plausible reaction pathway that considers the water corrosion process of the Schreibersite meteoric material operated by water micro solvation, carried out with periodic simulations at PBE-D*0 theory level. The main steps of the process involve the deprotonation of physisorbed water molecules at metallic sites and the following migration of the -OH moiety towards P. This process, reiterated several times, leads to more and more oxidized forms of P, until the highest possible oxidation state: $H_3PO_4$, the one of interest for prebiotic purposes. The modelled reaction pathway shows that the overall process is both thermodynamically and kinetically favored at the prebiotic conditions. We therefore predict, by means of quantum mechanical calculations, the feasibility of the water corrosion process of Schreibersite at temperatures compatible with those of the primordial Earth for the production of phosphates, a key ingredient for the emergence of life. Finally, we remind that the schreibersite surfaces are not behaving as a catalyst as they corrode by the water reactions. This means that the surface is continuously changing during the corrosion process and, accordingly, the perfect crystalline (001) surface should be intended as an exploratory model preferred to the most stable (110) surface for its higher reactivity. Nevertheless, this model is certainly not enough to fully characterize the complex reactivity of real nanoparticles of Schreibersite where a set of structural and imperfections (edges, kinks, vacancies, atom impurities, etc.) may profoundly affect the reactivity. To stay on the safer side, we stress that the present kinetic results should be considered as an upper limit estimate of the reaction barriers while future plans are in place to model corrosion processes on nanoclusters.

## Computational details

The water corrosion process of Schreibersite was studied by means of periodic DFT calculations carried out with the Vienna Ab-initio Simulation Package (VASP) code,[27–30] which uses projector-augmented wave (PAW) pseudopotentials[31] to describe the ionic cores and a plane wave basis set for the valence electrons. Geometry optimizations and frequency calculations were performed with the gradient corrected PBE functional,[32] with a posteriori Grimme D2 correction,[33] modified for solids (D*).[34] Moreover, C6 atomic coefficients related to polarizabilities on Fe and Ni metal atoms were set to 0 (*i.e.* no dispersion interaction contribution from metal atoms). On H, C and O atoms the original D* parameters were used. This method of choice is referred to as PBE-D*0 along the work. This setup was chosen according to the best results obtained in our previous work on the bulk and bare surfaces of schreibersite.[15] In the present case we are dealing with water reactivity, on an otherwise metallic system. Literature proved that pure GGA functionals overestimate reaction energies and underestimate the kinetic barriers due to delocalization error, larger for TS than reactants and products. Exact exchange contribution as in hybrid functionals mitigates, but does not eliminate, this problem.[36,37] In particular, careful studies have been applied to the water molecule, showing an error of the DFT in describing the OH bond breaking of about 3-4 kJ/mol.[35] Nevertheless, it is well-known since many years that the including of Hartree-Fock exchange leads to unphysical density of states and, accordingly, a bad description of the metallic character of transition metal systems.[38–41] Moreover, a recent paper pointed out that reaction kinetics occurring on transition metal surfaces, also involving OH bonds rupture/formation, do not follow the systematic trend reported for the gas-phase reactions, showing that in many cases pure GGA functionals (like PBE here adopted) outperform methods in higher steps of the Jacob's Ladder.[42] The cutoff energy of plane waves (which control the accuracy of the calculations) was set to 500 eV. The self-consistent field (SCF) iterative procedure was converged to a tolerance in total energy of $\Delta E = 10^{-5}$ eV for geometry optimizations, while for frequency calculations the tolerance was tighten to $\Delta E = 10^{-6}$ eV. For transition state search the NEB method was used,[43–47] and the structures were then refined with the DIMER method.[48–51] The tolerance on gradients during the optimization procedure was set to 0.01 eV/Å for each atom in each direction. The Monkhorst-Pack sampling of the Brillouin zone was used for the *k*-points mesh. Shrinking factors for the (001)2, i.e. the most stable among the different (001) terminations, (a = b = 8.984 Å, γ = 90°, see Figure S1 and Table S1 for details) surface have been set to (4 4 1). As VASP relies on plane waves basis set and, accordingly, surfaces are replicated also along the non-periodic direction, the vacuum space among fictitious replicas was set to at least 20 Å to minimize the interactions among replica images. Therefore, the final *c* cell axis was set to 40 Å. Geometry relaxations were carried out by moving all atoms in the unit cell while keeping the cell parameters fixed at the geometry optimized for the bulk structure to enforce the rigidity due to the underneath macroscopic surface. Vibrational frequencies were computed at Γ point, using the Phonopy code,[52] by numerical differentiation of the analytical first derivatives, using the central difference formula (*i.e.* two displacements of 0.02 Å for each atom in each (x, y, z) direction), in order to confirm that the optimized structure is either a minimum (all real frequencies) or a transition state (all but one real frequencies). Only O and H atoms were set free to vibrate. In the supporting information we show that the partial *vs* full calculation of the hessian matrix produces an error on the vibrational frequencies of at most 2.8%, and on the relative zero-point-energy and Gibbs free energy corrections of 2-3 kJ/mol, below the chemical accuracy of quantum mechanical simulations.

The kinetic rate constant *k* of water dissociation, was computed using the classical Eyring's transition state theory equation assuming a unimolecular process:[53,54]

$$k = \frac{k_B T}{h} e^{\frac{\Delta G^{\ddagger}}{RT}} \qquad 1$$

where $k_B$ is the Boltzmann constant, $T$ is the absolute temperature, $R$ is the ideal gas constant, $h$ is the Planck constant and $\Delta G^{\ddagger}$ is the difference of Gibbs energy between the transition state and the linked corresponding minimum. The half-life time $t_{1/2}$ has been estimated assuming first-order kinetics, as:

$$t_{1/2} = \frac{ln2}{k} \qquad 2$$

Visualization and manipulation of the structures and figures rendering have been done with the MOLDRAW[55] and POVRAY[56] programs.

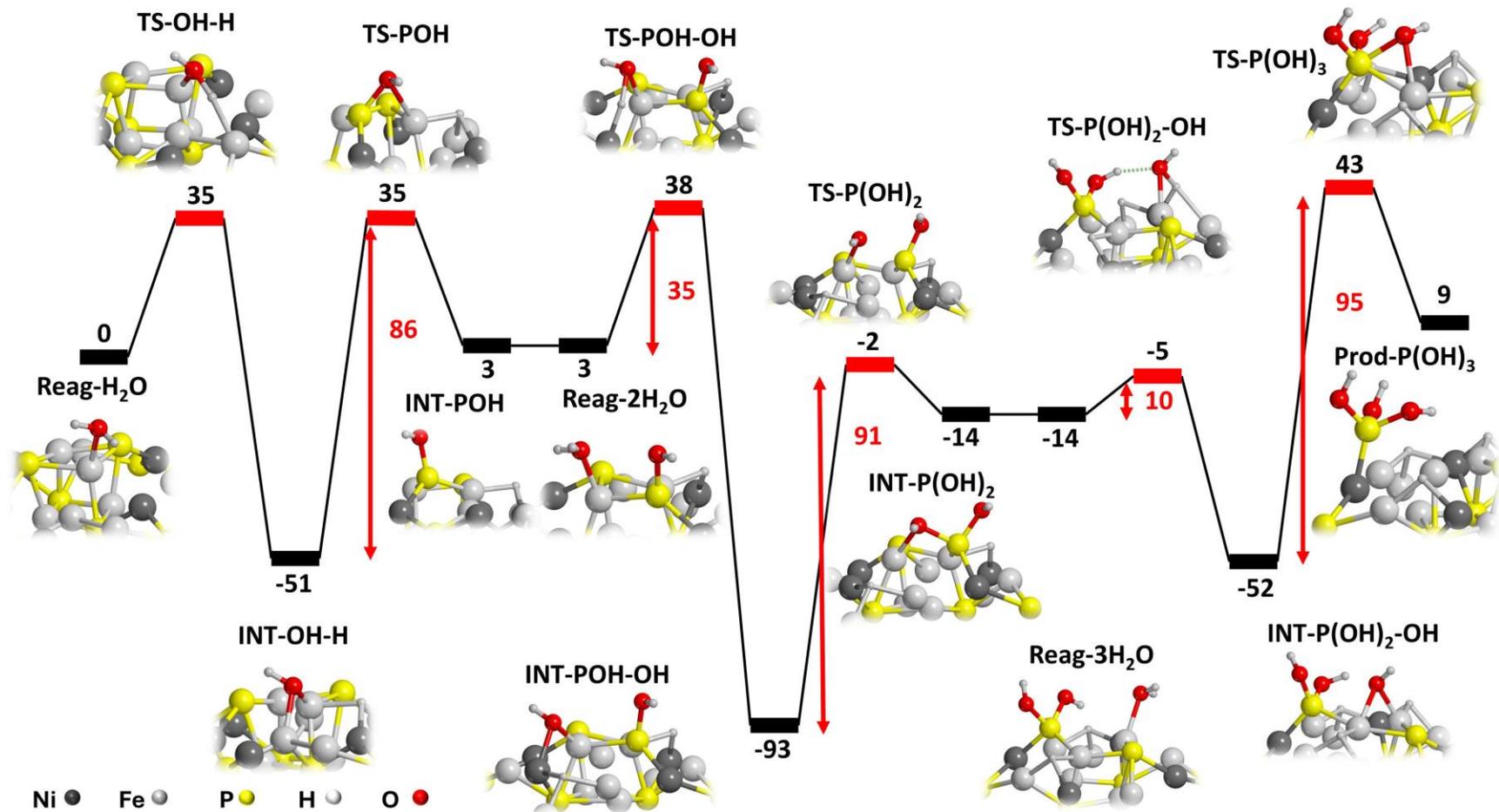

**Figure 1.** PBE-D*0-Gibbs energy reaction pathway at 298 K (in kJ/mol) of the water corrosion process on the (001)2 Schreibersite surface from 1 up to 3 water molecules. Atom colour legend: H in white, O in red, P in yellow, Fe in light grey, Ni in dark grey. Transitions states are highlighted in red.

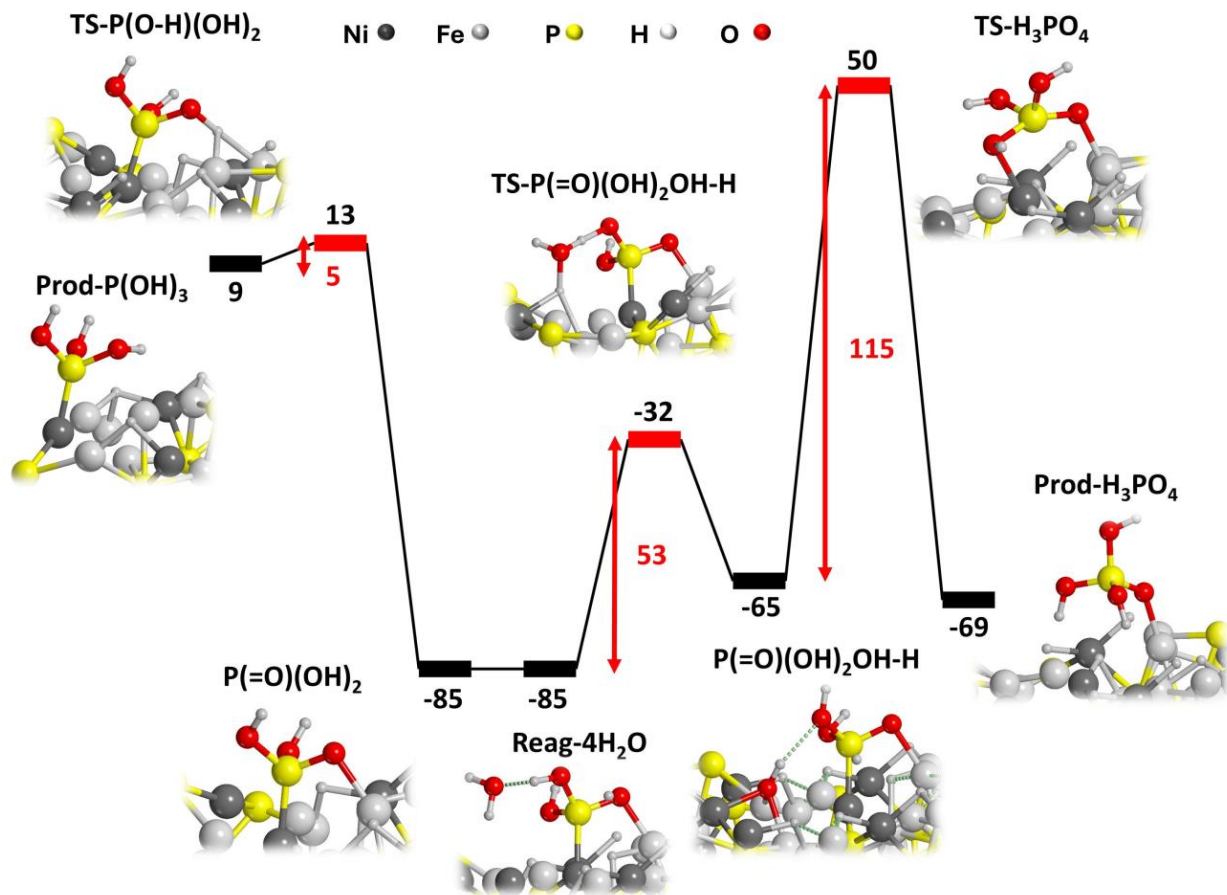

**Figure 2.** PBE-D*0-Gibbs energy reaction pathway at 298 K (in kJ/mol) of the evolution from phosphorous acid (P(OH)$_3$) to phosphoric acid (H$_3$PO$_4$) on the (001)**2** Schreibersite surface. Atom colour legend: H in white, O in red, P in yellow, Fe in light grey, Ni in dark grey. Transitions states are highlighted in red.

## ASSOCIATED CONTENT

**Supporting Information**. Schreibersite (001) surface structures and energies. Vibrational frequencies benchmark: full vs partial hessian. ΔE and ΔG calculated at various temperatures. All the optimized structures are included in an external .zip file in POSCAR format.

## AUTHOR INFORMATION

The authors declare no competing financial interests.

## FUNDING SOURCES

This work has been partially supported by the Spoke 7 "Materials and Molecular Sciences" of ICSC – Centro Nazionale di Ricerca in High-Performance Computing, Big Data and Quantum Computing, funded by European Union – NextGenerationEU, from the Italian MUR (PRIN 2020, "Astrochemistry beyond the second period elements", Prot. 2020AFB3FX) is gratefully acknowledged and by the Italian Space Agency (Bando ASI Prot. n. DC-DSR-UVS-2022-231, Grant no. 2023-10-U.0 "Modeling Chemical Complexity: all'Origine di questa e di altre Vite per una visione aggiornata delle missioni spaziali (MIGLIORA)" Codice Unico di Progetto (CUP) F83C23000800005). This project received funding from the European Research Council (ERC) under the European Union's Horizon 2020 Research and Innovation Program (Grant Agreement 865657) for Project "Quantum Chemistry on Interstellar Grains" (QUANTUMGRAIN). MICIN is also acknowledged for financing the projects PID2021-126427NBI00 and CNS2023-144902.

## ACKNOWLEDGMENT

We acknowledge the EuroHPC Joint Undertaking for awarding this project access to the EuroHPC supercomputer LUMI, hosted by CSC (Finland) and the LUMI consortium through the EuroHPC.

Regular Access call n° EHPC-REG-2023R03-117, and the CINECA award under the ISCRA initiative, for the availability of high-performance computing resources and support.

The authors acknowledge support from the Project CH4.0 under the MUR program "Dipartimenti di Eccellenza 2023–2027" (CUP: D13C22003520001).

For TOC Only

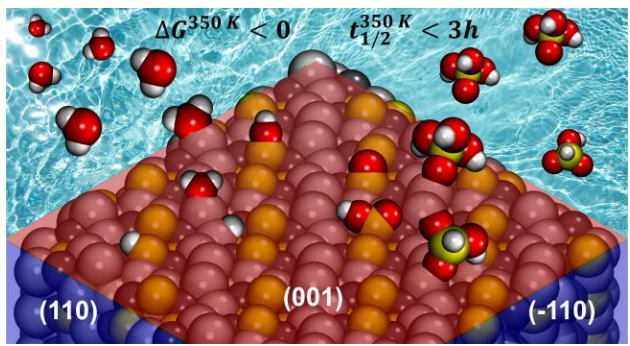

All parts of the TOC graphic are entirely original, unpublished, and created by one of the coauthors.